\documentclass[11pt]{article}
\usepackage{graphicx} 
\usepackage[margin=1in]{geometry}
\usepackage{natbib}
\usepackage{amsmath}

\title{Predicting Patient No-Shows in Community Health Clinics:\\A Case Study in Designing a Data Analytic Product}

\author{Roger D. Peng\\\textit{Department of Statistics and Data Sciences, University of Texas, Austin}}
\date{}

\begin{document}

\maketitle

\begin{abstract}
The data science revolution has highlighted the varying roles that data analytic products can play in a different industries and applications. There has been particular interest in using analytic products coupled with algorithmic prediction models to aid in human decision-making. However, detailed descriptions of the decision-making process that leads to the design and development of analytic products are lacking in the statistical literature, making it difficult to accumulate a body of knowledge where students interested in the field of data science may look to learn about this process. In this paper, we present a case study describing the development of an analytic product for predicting whether patients will show up for scheduled appointments at a community health clinic. We consider the stakeholders involved and their interests, along with the real-world analytical and technical trade-offs involved in developing and deploying the product. Our goal here is to highlight the decisions made and evaluate them in the context of possible alternatives. We find that although this case study has some unique characteristics, there are lessons to be learned that could translate to other settings and applications.

\vspace{1em}\noindent Key Words: data science, product development, dashboard, data analysis
\end{abstract}

\section{Introduction}
\label{sec:introduction}

The use of data science algorithms and products has proliferated across a variety of industries in recent years. The use of algorithms for decision-making purposes holds the promise of allowing for the optimization of scarce resources at an organization. For example, health care clinics that have appointments where patients fail to appear (i.e. no-show appointments) can strategically overbook appointment slots to maximize the productivity of the clinic. Such a practice is common in other industries such as with commercial airline travel~\citep{suzuki2002empirical}.

Building a data science product to aid in decision-making requires assembling available data, specifying the requirements for the output, developing and training a model, and deploying the model in a production environment. Detailed descriptions of how this process works and the trade offs that are made are generally lacking in the academic literature. Such work tends to happen inside private organizations that are not particularly incentivized to disseminate the details of their experience with product development. Case studies of statistical analyses and the application of specific statistical models are available, but these studies often focus on the modeling aspects and results while skipping over many of the analytical design decisions that can be critical to the success of a data science project. Having a detailed study of building a data science product would be valuable for training purposes and would allow data scientists to learn lessons from others' experience and to avoid repeating mistakes. In addition, having a description of how a data science product is built would allow for the decision-making process to be scrutinized and possibly inspire alternative approaches.

In this paper we describe the process of building a data science product used to predict no-show appointments in a health care clinic. In our description, we attempt to highlight the challenges that had to be addressed and the variety of strategies that were considered. In particular, some questions we address are:
\begin{itemize}
\item
Who are the stakeholders in this data science problem and what are their interests and incentives?
\item
What kinds of metrics are valuable in a prediction problem for this kind of setting? What metric can best aid in decision-making, particularly by a non-quantitative audience?
\item
How does a data scientist architect a data science product in an efficient manner?
\item
What is the nature of a generalizable solution in data science? Can data science work be sold as a product?
\end{itemize}
This case study describes the development of a product by Streamline Data Science, a data science technology company based in Baltimore, Maryland. The case study was developed using publicly available information provided by Streamline Data Science and the St.~Thomas Community Health Center. Additional information was obtained through interviews with Charlie Drain, CEO of Streamline, Dr.~John Muschelli, CTO of Streamline, and Dr.~Jeffrey Leek, a Co-Founder of Streamline.

\section{Background}
\label{background}

\subsection{The Patient No-Show
Problem}\label{the-patient-no-show-problem}

Medical clinics typically schedule appointments to see patients
throughout the day. However, for various reasons some fraction of those
patients do not show up for their appointments. These ``patient
no-shows'' are common throughout the industry. One
systematic review found that the mean rate of missed appointments was 15.2\%, with a median of 12.9\%~\citep{parsons2021patients}. Huang and Hanauer (2014) note that studies of patient no-shows have reported no-show rates ranging from 10\% to 30\%, although there can be variation attributable to the clinic, location, and provider specialty~\citep{huang2014patient}. 

Patient no-shows raise two immediate issues. The first is that the scheduled patient does not receive the care they need and the second is that patients who need care may be denied an appointment that day because there are thought to be no available appointment slots. In either case, care is not given to someone who needs it. If a patient is a no-show (or cancels near the time of appointment), it may not be possible to re-book the appointment in time. This leads to a third problem, which is under-utilization of clinic resources and provider time.

Overbooking appointment slots by scheduling multiple patients for the same time period is one strategy for addressing patient no-shows. While overbooking does not address the problem of the scheduled patient missing needed care, it does allow for other patients to receive care in a timely manner by filling in the no-show time slots. However, there is risk to arbitrarily overbooking appointment slots because if two patients show up for the same scheduled time, this can cause delays in the clinic and increase the general level of stress amongst providers (and patients). Therefore, clinics generally aim to implement an overbooking strategy that maximizes the number of patients it can see in a day (and thus increasing the clinic's overall utilization) while minimizing the chance that multiple people show up for the same time slot, thus causing delays and stress.

\subsection{St.~Thomas Community Health
Center}\label{st.-thomas-community-health-center}

The St.~Thomas Community Health Center (CHC) is a federally qualified
health center that provides primary, preventive, and specialty care to under-served populations in New Orleans, Louisiana. The CHC started in 1987 providing primary care services in the St.~Thomas Housing Development and since then has expanded to eight locations in the New Orleans area. As a federally qualified health center, St.~Thomas CHC is committed to providing health care to people regardless of the their ability to pay or their health insurance status~\citep{stthomaschc}.

St.~Thomas, like many clinics, had experienced consistent patient
no-shows. Alexandria Fischer, a senior data analyst at St.~Thomas, estimated that the patient no-show rate was about 25\% of appointments~\citep{fqhcconnect2022}. This rate was somewhat above average based on systematic reviews of patient no-show rates~\citep{huang2014patient}. Essentially, the clinic was running at 75\% utilization, which, from a financial and health care delivery perspective, was not optimal in the long-run.

Clinics like St.~Thomas had generally employed two broad categories of approaches to addressing the patient no-show problem.
\begin{enumerate}
\item
  Intervene to change patient behavior in order to decrease the no-show rate. Such interventions included call-backs and text reminders to patients in the days before the scheduled appointment.
\item
  Overbook appointments to allow for care to be provided to other patients during no-show appointments.
\end{enumerate}
These approaches were not mutually exclusive and St.~Thomas employed both of them to some degree. Fischer noted that the clinic felt that they were limited in their ability to change patient behavior because they did not necessarily have control over the factors that might cause patients to no-show (e.g.~conflicts with patients' work schedules, lack of transportation, weather). Therefore, they thought optimizing the overbooking process could improve the productivity of providers at the clinic while also increasing the overall amount of care that could be offered to patients each day.

\subsection{Streamline Data Science}
\label{streamline-data-science}

Streamline Data Science was founded in 2021 as a company that would help other companies with common data science challenges~\citep{streamline2023}. The co-founders included former professors from the Biostatistics department at Johns Hopkins University who had spent years dealing with messy and complex data problems in the health sciences. One co-founder, Dr. Jeffrey Leek, had previously started a company called Problem Forward Data Science that focused on providing fractional data science services to a variety of businessess. Speaking of his experience with Problem Forward, Leek wrote in a 2021 blog post~\citep{leek2021simplystatistics},
\begin{quote}
We were asked to do a lot of different types of data work, everything
from turning spreadsheets into dashboards to building complicated
forecasting models. But no matter the project, whether in government,
academia, or industry, we always ended up with the same problem.
\emph{We needed to clean the data before we could do the data science}.
\end{quote}
Sensing an opportunity, Leek eventually shuttered Problem Forward Data Science and soon after started Streamline Data Science in order to provide this exact service---tidying messy data for other people. With tidy data~\citep{wickham2014tidy}, companies could do the kinds of predictive analytics that they claimed they wanted to do. From the same blog post,

\begin{quote}
Customers wanted complex AI {[}artificial intelligence{]}, insightful
dashboards, or easy reports, but the data just weren't ready for that
yet. And we wasted a huge amount of time cleaning the data over and over
again. We realized that the most common challenge companies have is that
their data processing and management pipelines aren't ready for
analytics.
\end{quote}

The value proposition of Streamline was predicated on the idea that data
often get ``stuck'' because most companies' software systems are a mix
of systems that fundamentally do not work well together. Furthermore,
there is typically no real incentive for one software vendor to make
their systems compatible with another vendor's system. But predictive
models, dashboards, and other analytic summaries derive value from the
fact that they integrate data across the company's various systems. Such
integration is not possible unless data can be extracted, cleaned, and
merged together. For example, predicting whether a patient will show up
for an appointment at a clinic may require combining data from an
electronic health record and the clinic's scheduling software. This is
where Streamline would come in, providing ``tidy data as a service''. In a
sense, Streamline was offering to ``liberate'' and clean data that was
stuck in various places, whether it be Excel spreadsheets, proprietary
software systems, or other manual processes.

From a business model perspective, Streamline figured that once a company saw the benefits of having their data ``tidy-ed'', and thus amenable to all kinds of predictive modeling and analytics, they would be inclined to pay Streamline to manage this process over time. Streamline would also offer to build predictive models or dashboards for companies that employed Streamline's data cleaning services. Because the underlying software systems generating the data were unlikely to get much easier to use from a data perspective, companies would always need the services offered by Streamline to monitor the system's data and ensure that the data pipeline was generating the proper results.

Leek described one data pipeline that Streamline built early on (from the 2021 blog post):

\begin{quote}
They needed someone who could come in and set up a data processing
pipeline for them, manage it, and make sure the data were up to date.
Some people call this Extract Load Transform (ELT), but we found it goes
a bit beyond that. It is figuring out what format is most useful for the
people who rely on data and working backward to create a customized and
unique data pipeline that gets the data ready to use. The ELT pipeline
we set up is designed to consistently output ``tidy data'' that makes it
easy for our customers to use {[}business intelligence{]} tools like
Tableau or Looker and to ingest their data without having to do all the
ugly data work that is painful and time-consuming.
\end{quote}
Streamline recognized that tidying data was not a purely mechanical
or technical process. Rather, design thinking was required in order to determine the
use cases that people had for the data and to understand the business
problems they were ultimately trying to solve. Only then could the data be tidied for the appropriate analysis.

As a young startup, Streamline faced two key questions involving both its business model and its technical strategy. A critical business model question for Streamline was whether data cleaning---tidy data as a service---was something people would be willing to pay for. Would customers appreciate the effort and time that went into tidying data and the value of having tidy data for building analytic models and products? If so, how big was this market? From a technical strategy perspective, a key question was whether tidying data was a ``productizable" service, i.e. something that could be sold in a repeatable manner. Experience had shown that every company had somewhat different data problems and questions, making it difficult to apply the same tidying process across different customers. Could Streamline engineer a data tidying ``product" that could be leveraged consistently for every customer or would they have to start from scratch with every engagement, significantly increasing their development costs?

\section{Building an Analytic Product to Predict Patient No-Shows}
\label{predicting-patient-no-shows}

St.~Thomas approached Streamline Data Science to look at the overbooking problem and associated data tidying issues with the hopes that Streamline could make the entire process more efficient. Streamline CEO Charlie Drain noted that St.~Thomas, like many organizations, did not have a large data science staff, and would have faced difficulty devoting significant time and resources to building out the system they envisioned~\citep{draininterview2023}. St.~Thomas initially considered handling the overbooking problem by looking at average historical no-show probabilities for each patient. The idea would be that each day, a spreadsheet of each provider's scheduled appointments would be emailed to the providers by the staff. Included with the patient information would be the patient's historical no-show rate, which for a given patient was
$$
\text{No-show Rate}=\frac{\text{Total \# of missed visits}}{\text{Total \# scheduled appointments}}.
$$ 
Providers could then decide for themselves which appointment slots to overbook. 

According to CEO Drain, the initial vision for this project was largely manual and therefore less efficient than it could be. Furthermore, the determinations for overbooking were based on a single summary statistic, the long-term no-show rate. Drain felt that Streamline could make improvements to St.~Thomas's approach in two key areas. The first area was automating the tedious and manual data merging process that was required to compute the daily no-show rates for each of the scheduled appointments. The second area was leveraging information from additional clinic and patient covariates to improve the predictions of patient no-shows using machine learning. Streamline felt there was substantial information that could be used in the clinic's existing data systems to improve prediction performance. Ultimately, Streamline built a pilot product using some historical data in order to demonstrate their capabilities.

\subsection{Initial Design Challenges}

Dr.~John Muschelli, co-founder and the Chief Technical Officer of Streamline, was charged with designing and building a solution for the patient no-show problem. Muschelli saw the problem as representing a combination of data, analytical, and software challenges~\citep{muschelliinterview2023}.

Getting the data from their electronic health record (EHR) was likely to be a challenge. EHR's, in general, had been designed for providers to input and store information about patients. They were not designed for doing research or experimentation and it was often difficult to extract data from them. CEO Drain noted ``I'm always astounded by how slow it is to get the data.'' In most cases, obtaining data from the EHR required highly specific programming to be done as there were typically no well-maintained or well-curated APIs for accessing the data from the EHR. Streamline worked closely with the clinic's EHR provider in order to establish a workflow to download the data.

Identifying a metric that could be used to make decisions about overbooking would require discussion and collaboration with the clinic providers and schedulers. The clinic had used each patient's historical rate of no-shows, which was essentially the probability of not showing up for an appointment. However, it was not necessarily clear that this was the best metric to use for this application. Was there something else that might be more useful or interpretable to the providers who were tasked with making the overbooking determinations?

Building a prediction model that could predict the agreed-upon metric with a reasonable amount of performance was another challenge. Muschelli needed to build a model that would suit the available data from the clinic and be efficient to run on cloud-based platforms. He also had to determine what performance metrics to optimize that would also represent well the requirements of the clinic. He noted that accuracy (i.e. the proportion of times the model made the correct prediction) was not necessarily an ideal metric because if the overall no-show rate was around 20\%, then he could guarantee a model with 80\% accuracy by simply predicting that every patient would show up for their appointment. Once these issues were addressed, Muschelli also had to write a software system that cleaned the clinic's data, implemented the prediction model, and provided an interface that clinic staff could use to interact with the prediction model. 

All of the challenges described above had implications for how the software would be designed. Streamline saw the patient no-show problem as something that would be common to a number of clinics, and so any solution they developed could potentially be sold to a range of customers. However, if the software was not reusable to a large extent, Muschelli and his team would need to re-engineer the software for each customer, making it difficult to scale the system rapidly to a large number of clinics.

\subsection{Pilot Data}
\label{pilot-data}

Because Streamline had decided to build a pilot version of the no-show prediction system, an early question for Muschelli involved deciding how to build the demonstration without requiring a transfer of significant amounts of protected health information (PHI) about patients to Streamline. Such a transfer would likely require a strict formal agreement between St.~Thomas and Streamline and would also necessitate higher levels of data security for Streamline. While such an agreement might be made further in the process, arranging the agreement would take a considerable amount of time and would hinder Streamline's ability to demonstrate a prototype. Muschelli thought that, at least in the early stages, it would be best if they did not use any PHI from patients and rather saw how good of a model they could build without it. Initially, Muschelli obtained the following data on each appointment in
the clinic:
\begin{itemize}
\item
  An identifier for the clinic provider
\item
  The provider's specialty
\item
  Appointment date and time
\item
  An (anonymous) identifier for the patient
\item
  An indicator of whether the patient showed up
\item
  The location of the scheduled visit
\item
  The date/time that the appointment was made.
\end{itemize}
From this data, some amount of feature engineering could be done to compute additional quantities like the time of day, season of the year, day of week of the appointment, and the lead time for the appointment (i.e. the amount of time between when the appointment was made and the time of the appointment itself).

\subsection{Analytic Strategy}
\label{sec:analytic-strategy}

For the prediction model for patient no-shows, Muschelli was confronted
with a question of how to present the information to the clinic staff and providers. The primary aim was to present some sort of summary regarding the likelihood of a patient showing up for a scheduled appointment. 

One option was to use a table in a spreadsheet to present the appointment information and no-show probabilities each day. The advantage of using a table like this included integrating with tools familiar to clinic staff and providers, such as Microsoft Excel, and being able to include a dense amount of patient information (e.g. patient name, reason for appointment, etc.) in addition to the likelihood of showing up for the appointment. One downside of a table was that it would require providers to interpret numerical probabilities and translate those probabilities into a decision about overbooking an appointment. 

Another option was to build a graphical representation of the no-show probabilities that could indicate where in the schedule overbooking opportunities might exist. This type of display could be more eye-catching and perhaps interpretable by using a color-coded scheme to indicate which appointment slots might be likely to be missed (and therefore could be overbooked). While less information about each individual appointment/patient could be presented in the graphical display, there was the possibility of displaying more information from later in the week (rather than just a single day's worth of appointments) in order to get a more global picture of the clinic schedule.

The question of what metric to display was resolved by choosing the \textit{expected number of missed appointments per hour-long block of appointments}. This outcome metric represented a slight shift in perspective from thinking about the probability of an individual appointment being missed to thinking about the rate of missed appointments over a pre-specified block of time. For example, if an hour-long block of time had four 15-minute appointments scheduled, and each appointment had a 25\% chance of being a no-show, then the expected number of missed appointments in that hour-long block was $0.25 + 0.25 + 0.25 + 0.25 = 1$. In this situation, a provider might decide to overbook that hour block by one appointment, so that the expected number of appointments was still~4.s

With this outcome metric, the solution Muschelli developed was to present a graphical display of the provider-specific expected number of missed appointments for each hour-long block during the day. The approach would be to
\begin{enumerate}
\item
  Have the model predict the probability of a no-show for each scheduled
  patient and map the patient's appointment time to the corresponding hour-long block that day.
\item
  For each hour-long block on a given day and for a given provider, sum
  the predicted probabilities to obtain the expected number of
  missed appointments in that hour.
\item
  For each provider, produce a color-coded graphical display that would indicate whether the expected number of no-shows in a given hour-long block was less than 1 (yellow), between 1 and 2 (orange), and more than 2 (red).
\end{enumerate}
With this color coding scheme, Muschelli could present the summarized
data in a dashboard that could be updated in an automated fashion with
each day's scheduling and patient data. The format for presenting the data was a heat map with the days of the week on the x-axis and the 1-hour-long appointment blocks on the y-axis. 

\begin{figure}[tbh]
\centering 
\includegraphics[width=4in,height=4in]{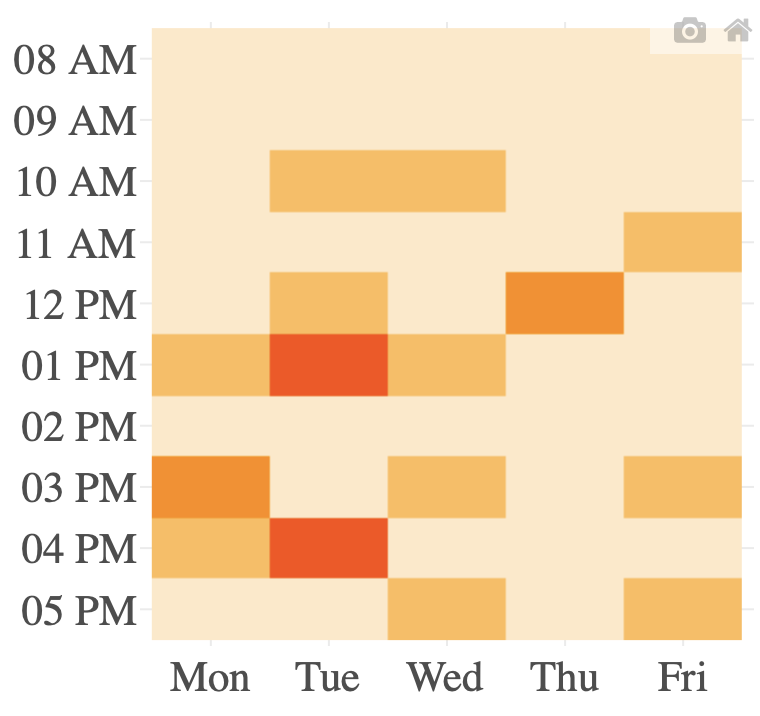}
\caption{Example heat map of expected number of no-shows per hour-long appointment block~\citep{streamline2023}.}
\label{fig:heatmap}
\end{figure}

Figure~\ref{fig:heatmap} presents an example of what this heatmap would look like. In this example, the 1pm and 4pm blocks on Tuesday have an expected number of missed appointments that is greater than 2. Therefore, this particular provider might consider targeting those hour blocks for potential overbooking.
The heat map gave schedulers a quick look at the week's appointments
and highlighted areas where overbooking might be a possibility. Another
benefit was that because all staff at the clinic, including providers,
could access the heat map and see where the likely no-shows were to
appear, providers would have a better understanding of why patients
might be overbooked in certain time slots. The data provided a concrete
basis for having an informed discussion about scheduling and
overbooking.

As for the prediction model itself, Muschelli chose to build it using a random forest model while optimizing the area under the ROC curve. While numerous other models could have been used in this application, he felt random forest generally provided good performance and was reasonably time- and memory-efficient. Specifically, he used the \texttt{ranger} package~\citep{ranger2017} for its speed with large datasets and embedded it within the \texttt{tidymodels} framework~\citep{tidymodels2020}, which provided a well-designed structure for training and testing the model. The aim would be for this model to provide the probability of an individual patient being a no-show at their scheduled appointment time. This probability would later be translated into the expected number of missed appointments for presentation in the graphical display. Muschelli was able to use 2 years of historical data from the clinic to build the model. 

The dashboard was built using Shiny~\citep{shiny2023} and included other pieces of information in contextual pop-up windows (i.e. tool tips), including details of each appointment, individual patient no-show probabilities, and the ability to filter on specific providers, specialties, or time periods. Because the Shiny dashboard could be presented using a web browser, clinic staff were able to include it amongst the open tabs in their existing browser-based workflow.

\subsection{Technical Strategy}
\label{sec:technical-strategy}

As part of building the dashboard for St.~Thomas, Muschelli aimed to build a software architecture that would be replicable across applications and clients with the minimum amount of re-writing of code. Ultimately, there would always be some client-specific code that would need to be freshly written. But the idea was to isolate that code and modularize the design of the system so that implementing client-specific details did not also require significant changes to other parts of the software.

Muschelli decided from the start to build the software product using R~\citep{R:2023}. He felt the R ecosystem was sufficiently mature and diverse enough so that all of the necessary components were available. He then designed an architecture by writing four different R packages for implementing the final prediction model and dashboard.
\begin{enumerate}
\item
One package was dedicated to client-specific aspects of the project. Some technical information needed to be hard coded into the package such as web URLs, API endpoints, and other site-specific details. The other key tasks for this package was implementing client-specific data cleaning and mapping of variable names to a common schema.
\item
A second package implemented the dashboard itself as a Shiny app that could be run in the cloud and viewed using a web browser. The Shiny app was agnostic to the data coming in and did not need to know any real details about the client or about the prediction algorithm. Therefore, it made sense to implement the dashboard as a separate package. Once an estimated no-show probability was calculated, this package's job was to present it in the heatmap representation alongside other clinic and patient data that might be passed to it from other packages. 
\item
A third package was written to implement the no-show prediction algorithm. This package stored the trained and tested algorithm and accepted new inputs for predicting no-show probabilities. The model, once trained, was frozen in place and not updated with subsequent data.
\item
The final fourth package was the ``glue'' that ingested the data, executed any processing, fed the data to the model, and saved the final output that could be read by the Shiny app dashboard. This pipeline was implemented using the \texttt{targets} package~\citep{targets2021}, which simplified the execution of tasks that needed to be run in a specific order, as well as the management of their dependencies.
\end{enumerate}

Of the four R packages, the first package, which implemented
client-specific details was clearly not generalizable or re-usable
accross different customers. However, it was possible that the other three packages could be re-used with minimal modification or
customization. Different clients would likely want to customize the
dashboard implemented in the second package, but such customizations
would not likely require major changes to the underlying code. The
prediction model in the third package would need to be adapted to the
data and features available from a specific client, but the underlying
framework for training, testing, and deploying the model could be
preserved. The fourth package, which executes the entire data pipeline,
would hopefully not require much modification, as long as the interfaces
between the various modules (i.e.~raw data, prediction algorithm, Shiny
app) were well-specified and general enough to handle different
situations.

\section{Results}\label{results}

\begin{figure}[tbh]
\centering 
\includegraphics[width=6in]{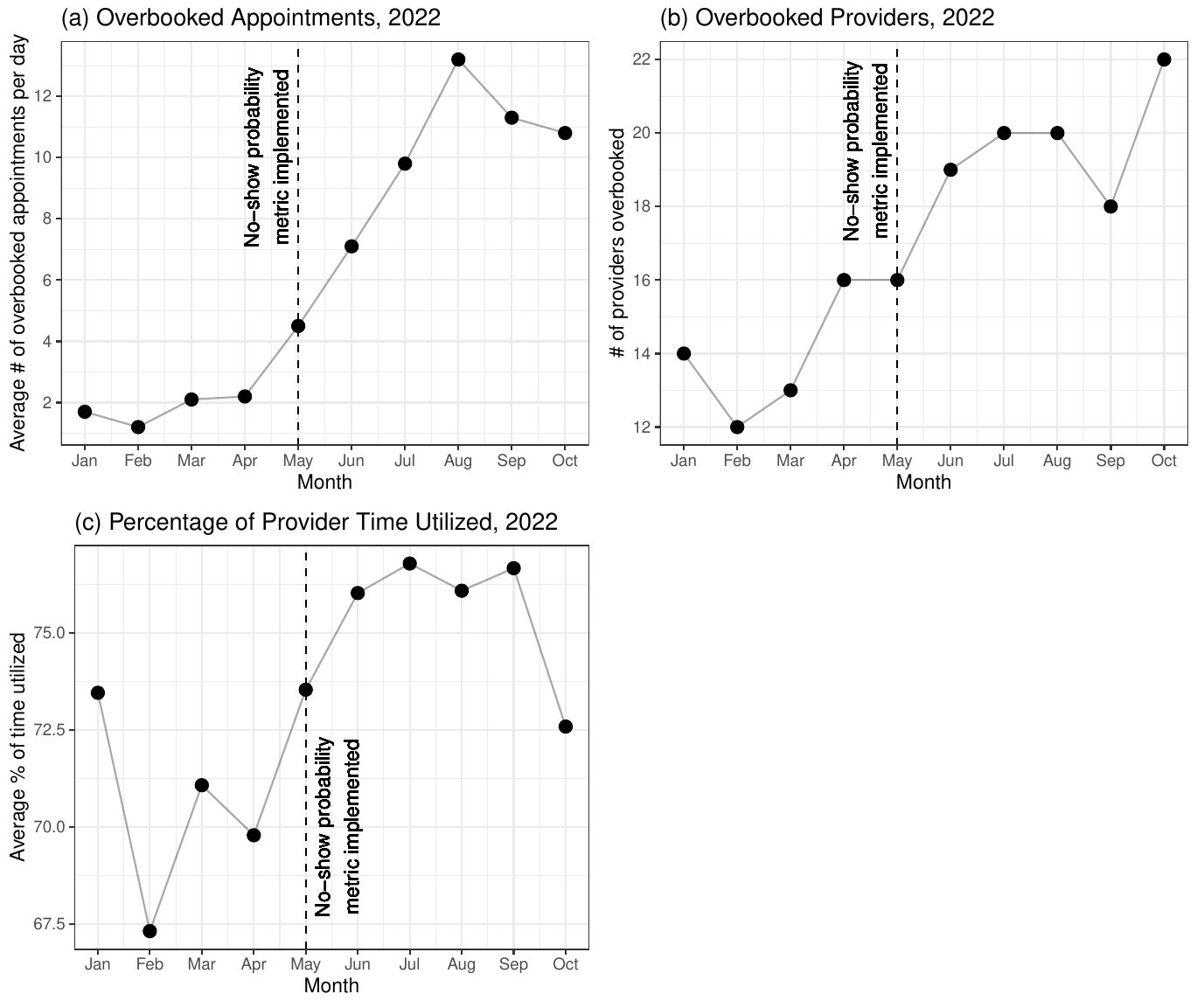}
\caption{Results at St.~Thomas CHC from initial implementation of patient no-show: (a) Average number of overbooked appointments per day; (b) Number of overbooked providers; and (c) Percentage of provider time utilized~\citep{fqhcconnect2022}}
\label{fig:overbookingresults}
\end{figure}

St.~Thomas piloted the patient no-show probability metric in May of 2022 in collaboration with Streamline. For this pilot, the full random forest machine learning model was not used, but much of the back end work done by Streamline for computing the no-show rate was being deployed. The preliminary results from the pilot are shown in Figures~\ref{fig:overbookingresults}(a--c); data were only available until October 2022. 

In the months before implementing the algorithm, St.~Thomas had generally overbooked between 1 and 2 appointments per day with a provider utilization rate ranging from 67\% to 73\%. After implementing the algorithm, Figure~\ref{fig:overbookingresults}(a) indicates that overbooking initially jumped to 4.5 overbooked appointments per day and then slowly increased to over 10 overbooked appointments per day in October 2022. Figure~\ref{fig:overbookingresults}(b) shows that the number of overbooked providers increased after the probability metric was implemented to between 18 and 22 doctors each month. In addition, Figure~\ref{fig:overbookingresults}(c) shows that provider utilization increased to a range of 72\% to 76\%.

While the early data were encouraging, with just a few months of
results, it was not yet clear if the changes were a temporary excursion or represented a more long-term phenomenon. Seasonal factors might be at play in this sample, as well as potentially other time-varying confounders. In particular, the data suggest that trends in each of the three outcomes were increasing before the implementation of the probability metric. Therefore, it's likely more data would be needed to draw a concrete conclusion regarding the effectiveness of the metric on these outcomes. 

\section{Discussion}
\label{discussion}

The goal of this case study was to describe the process of designing and building a data analytic product in a real-world setting. The product in this case was built for the purpose of predicting patient no-shows in a community health clinic. Our focus here was on the decisions made and the technical and statistical considerations around building the statistical software. While the success of the product is yet to be determined and will likely require more data to make any conclusions, we can still learn from the process of building the product and the decisions made by the people at Streamline Data Science. 

Individual case studies depicting data science problems are valuable in their detailed presentation of the process, but it sometimes can be difficult to extract generalizable concepts that are applicable to other situations in which the details might vary greatly. Ultimately, a question remains regarding what can be generalized from a sample size of one? However, if we take the narrative presented by this case study of Streamline's experience building a patient no-show prediction model and combine it with our collective experience in similar (if not identical) situations, we believe that some common themes do emerge. 

In this section we will revisit some of the key decisions made by Streamline and consider them in the context of possible alternatives. In particular, we will focus on
\begin{enumerate}
    \item The choice of the R programming language vs. other possible languages
    \item The outcome metric used for displaying the prediction results
    \item The design of the graphical display of the results
    \item The four-package structure of the product implementation
\end{enumerate}
These choices were made at various points in the product development timeline and we will explore the extent that these choices may have affected the direction of the product. We discuss them below in approximately the order in which each choice was made in the timeline of the product.

\subsection{Programming Language and Ecosystem}

At some early stage of a product development timeline a choice must be made regarding what programming language or languages are going to be used to implement the product. Streamline chose the R programming language for this product in part because it was most familiar to the team and because it had a robust ecosystem of packages for doing many of the tasks they would need. Muschelli noted that there were now R packages for interfacing with Google Cloud Platform, Google Big Query, and Amazon Web Services, doing authentication of users, training and deploying machine learning models, and managing a complex data pipelines. 

Streamline's reliance on the extensive R package ecosystem and R's ability to interact with a range of outside services likely limited the potential alternative programming languages from which to choose. Muschelli was additionally intrigued by the recently released \texttt{tidymodels} series of packages which enabled rapid development of machine learning models. The team could have chosen Python, which has a similarly vast ecosystem of packages, but it likely would not have made much difference in the final product given the significant overlap between the capabilities of the two languages. Furthermore, it was unlikely that Streamline would be able to take advantage of many of Python's strengths. For example, although Python excels in its implementation of various deep learning frameworks, those approaches were not used for this product.

\subsection{Software Structure of Product Implementation}

It seems clear that the four-package structure described in Section~\ref{sec:technical-strategy} was designed with multiple separate goals in mind. The primary goal, of course, was building a product that St. Thomas could use to solve their problem of predicting patient no-shows. However, the solution to that problem did not inherently demand the multi-package structure built by Streamline. Rather, a monolithic structure, with all aspects of the solution built into a single package, might have sufficed. Indeed, sometimes efficiencies can be gained by not separating out aspects of a larger program into separate modules.

A critical issue for Streamline was whether the solution they built for St.~Thomas could be generalized to other clinics that might operate in different ways or use different kinds of data. If the company was going to grow successfully, it would need increase the number of deployments of its dashboard while minimizing the need to build a custom solution every time. With an eye toward future business, a second goal for Streamline was maximizing the flexibility of the product while minimizing the need for custom programming at each deployment. CEO Drain thought that the front end of the product--the user-facing dashboard--could be re-used because activities like scheduling and clearing the patient wait list were very similar across clinics. However, the backend parts of they system, like the data processing and the prediction model, might need to be tuned for each new client. Given the small size of Streamline, it was important to minimize re-writing code from scratch so that they could build products for clients quickly. However, the growing company had only deployed the dashboard in a few places and was still gaining experience in how to implement these systems at a larger scale.

A final, perhaps secondary, goal was related to minimizing the maintenance costs of the overall system. A common downside of modular systems is their increased complexity over monolithic systems. Muschelli's use of the \texttt{targets} package allowed him to manage the complexity of the various packages and their multitude of dependencies. The modular nature of the package system allowed for individual pieces of the product to be modified or improved without affecting another part of the product.

\subsection{Outcome Metric}

The decision over the outcome metric that would be used for reporting reflected an early difference between what St.~Thomas had previously conceived and what Streamline ultimately chose to implement. St.~Thomas had originally consider displaying for a given patient the historical rate of now-shows and allowing the individual providers to interpret that with respect to overbooking. Streamline chose to display the expected number of no-shows per hour-long block of appointments. 

Muschelli thought this particular type of metric allowed for more flexibility for the providers and the clinic by allowing them to choose a threshold for overbooking based on a specific (expected) number of no-shows in an hour-long appointment block. He felt this was better than simply categorizing each appointment as a definite ``no-show" or not, which would rely heavily on the model's prediction performance. By aggregating over hour long blocks, they could effectively smooth over some of the model's uncertainty in predicting no-shows. This approach essentially answers the question of how many ``effective appointments" there are in this appointment block as opposed to how many appointments are \textit{scheduled} in this block.

An alternative that Streamline could have taken, for example, would have been to compute the probability of "at least one no-show" in an hour-long appointment block. Such an approach would present a probability metric instead of an expected number but would similarly aggregate across the predictions in an appointment block. That said, it is not clear if this would be more or less interpretable to the providers that would need to make decisions based on the metric.

\subsection{Presentation of Results for Decision-Making}

Given the chosen outcome metric to report, there was still the question of how to present the prediction results to those who would need to make decisions about overbooking (in this case, the individual St.~Thomas providers). Streamline chose to use a heatmap type of presentation that would color-code the expected number of patient no-shows in each hour-long appointment blocks. In this display, the darker red colors would highlight the blocks with the higher expected no-shows. While this approach did not allow for the presentation of detailed appointment-level information, Streamline overcame this limitation by using what Muschelli characterized as ``lengthy" tool tips, or contextual pop-up windows, containing that information. Essentially, they took advantage of the multi-dimensionality of the graphical user interface to surface information at the moment when it might be needed.

Rather than develop a new graphical presentation system, Streamline could have made their predictions and then tried to find a way to integrate into the existing approach of presenting information in a spreadsheet format. This approach likely would have minimized friction for the clinic staff with respect to adopting a new tool. However, Streamline felt the graphical presentation would provide useful additional information to the clinic staff, such as the multi-day format. Furthermore, there would be minimal overhead in learning a new system given that the staff was already using a variety of web-based tool like the Shiny dashboard. That said, Streamline's approach would require the clinic to learn yet another tool, which, if not properly integrated into the clinic's workflow, had the potential to be ignored.

\section{Summary}

This case study of Streamline Data Science building an analytic product to predict patient no-shows exhibits a few themes that may be translatable to other data science problems. The first theme was related to the need for Streamline and St.~Thomas to align themselves regarding what kind of product St.~Thomas needed and what Streamline was capable of building. St.~Thomas's original conception of the solution (providing the historical rate of no-shows) was a reasonable approach based on their previous experience. However, Streamline was able to bring to the discussion a wider range of approaches involving machine learning models and graphical presentations. From the data scientist's perspective, it can be tempting to merely insist on a technically superior solution and to ignore the needs of the product's consumer. However, such an approach risks providing a product that is likewise ignored in turn and therefore does not actually solve the problem at hand. Utlimately, Streamline and the team at St.~Thomas worked well together and were able to agree on a design.

The second theme that emerged was the need for building a generalizable solution. In many situations where questions need to be answered with data, data scientists are confronted with the question of whether to ``solve this problem" or to ``solve problems like this problem." For example, scientists may need to choose between writing some code to clean a dataset or to write a software package that can clean future data from the same data generating mechanism. The desire for generalization in data analysis is a frequent cause of tension as it can conflict with the need to solve a specific problem quickly. Streamline attempted to address this conflict by modularizing their software implementation in a manner that would allow them to reuse code in future projects.

It was clear that each project that Streamline took on had a consultative aspect where the company learned about the client's problems and needs, as well as a technological aspect where Streamline handled the client's data. While it might not be possible to eliminate the consultative aspect, as every client is a little different, the technological aspects potentially could be modularized and rapidly adapted to each new client's data. Such an approach, if successful, would validate the company's goal building a tidy data ``product'' that could help a wide variety of customers.

\bibliography{refs}
\bibliographystyle{agsm}

\end{document}